%% file: main.tex
\documentclass[10pt]{article}

\usepackage{fullpage,parskip}

\usepackage[utf8]{inputenc} 
\usepackage[T1]{fontenc}
\usepackage{url}
\usepackage{ifthen,color,dsfont}
\usepackage{cite,comment}
\usepackage[cmex10]{amsmath} 

\usepackage{mathrsfs}
\usepackage{bbm}
\usepackage{amssymb}
\usepackage{amsthm} 
\newtheoremstyle{sltheorem}
                {}          
                {}          
                {\slshape}  
                {}          
                {\bfseries} 
                {.}         
                { }         
                {}          
\theoremstyle{sltheorem}

\newtheorem{theorem}{Theorem}
\newtheorem{remark}{Remark}
\newtheorem{lemma}{Lemma}
\newtheorem{corollary}{Corollary}

\DeclareMathOperator*{\argmin}{arg\,min}
\allowdisplaybreaks

\usepackage[pagebackref, hidelinks, colorlinks=true, citecolor=black!30!blue,linkcolor=black]{hyperref}
\usepackage[
noabbrev, 
capitalize
]{cleveref}
\usepackage[dvipsnames]{xcolor}
\usepackage{graphicx}

\input{defns}

\begin{document}
\title{Bayesian Despeckling of Structured Sources} 



\author{
    Ali Zafari$^*$, Shirin Jalali\thanks{Rutgers University, New Brunswick, Department of Electrical and Computer Engineering (ali.zafari@rutgers.edu, shirin.jalali@rutgers.edu).}
}
\date{}

\maketitle

\begin{abstract}

   Speckle noise is a fundamental challenge in coherent imaging systems, significantly degrading image quality. Over the past decades, numerous despeckling algorithms have been developed for applications such as Synthetic Aperture Radar (SAR) and digital holography. In this paper, we aim to establish a theoretically grounded approach to despeckling. We propose a method applicable to general structured stationary stochastic sources. We demonstrate the effectiveness of the proposed method on piecewise constant sources. Additionally, we theoretically derive a lower bound on the despeckling performance for such sources. The proposed depseckler applied to the 1-Markov structured sources achieves better reconstruction performance with no strong simplification of the ground truth signal model or speckle noise.
\end{abstract}

\section{Introduction}

\subsection{Problem statement}
Multiplicative noise, commonly referred to as \emph{speckle} noise, poses a significant challenge in coherent imaging systems such as synthetic aperture radar \cite{argenti2013tutorial}, optical coherence tomography \cite{bianco2018strategies}, and digital holography \cite{schmitt1999speckle}. The inherent non-linearity of the multiplicative noise model complicates the analysis and design of optimal despeckling algorithms—denoising methods tailored to address multiplicative noise and recover the underlying signal—even for relatively simple structured sources \cite{touzi2002review}.

Mathematically, the despeckling problem can be formulated as follows. Consider a stationary stochastic process ${\bf X} = \{X_i\}_{i \in \mathbb{N}}$, $X_i\in\Xc$, observed under a \emph{multiplicative noise model}, $Y_i = X_i W_i$, where $W_i$ represents the speckle noise.  In most coherent imaging applications, the speckle noise process is assumed to be fully developed and is therefore modeled as Gaussian \cite{goodman2007speckle}. Here, we assume that $\{W_i\}_{i}$ are independent and identically distributed as standard normal random variables, $\Nc(0,1)$. The despeckling (or denoising under speckle noise model) goal is to recover $X^n$ from speckle-corrupted measurements $Y^n$.

In a Bayesian framework, given the source distribution $p(x^n)$, the optimal MMSE despeckler lets $\Xh^n=\E[X^n | Y^n]$. However, even for additive noise, direct computation of $\E[X^n | Y^n]$ is often intractable and highly challenging. Additionally, the source distribution $p(x^n)$ is rarely accessible in practice, and typically, only samples from this distribution are available. These challenges raise a fundamental question: \\
\noindent{\bf Question.}
    Can we design a theoretically founded, computationally efficient framework for despeckling that applies to a broad class of structured sources?

\subsection{Related work}

Theoretically derived Bayesian despeckling methods can be broadly categorized into two approaches: adaptive linear minimum mean squared error (MMSE) filtering and maximum a posteriori (MAP) estimation \cite{touzi2002review, argenti2013tutorial}. In early work, \cite{lee1980digital} proposed an affine approximation to the nonlinear speckle model, minimizing the $\ell_2$ norm and matching first-order moments to derive the MMSE estimator. A fully driven linear MMSE approach was later introduced in \cite{kuan1985adaptive}, avoiding the approximation errors of the affine model. Both methods rely on local statistics calculated over pre-defined windows, yielding comparable despeckling performance in practice \cite{lopes1990adaptive}. A refinement for images was proposed in \cite{lee1981refined}, leveraging the local gradient to adaptively redefine neighborhoods for improved statistical estimation.

Beyond linear MMSE methods, MAP estimators have been extensively studied, often requiring strong assumptions about the signal prior distribution. For example, \cite{lopes1990maximum} considered a MAP framework with a gamma prior, in contrast to the Gaussian prior used in \cite{kuan1987adaptive}. These methods generally assume local stationarity of the signal and parameterize its distribution using moments computed within a local window.

Other approaches explore regularization-based techniques, such as total variation (TV) minimization \cite{rudin1992nonlinear}. For speckle noise, \cite{aubert2008variational} combines a gamma likelihood fidelity term with the TV regularizer, though the inherent nonconvexity of the speckle model limits its practical utility. A log-transform-based linearization of the speckle model was proposed in \cite{shi2008nonlinear}, where the additive noise is denoised and rescaled back by exponentiation. However, this approach suffers from degraded performance due to mismatches between the log and signal domains.

Additionally, drawing inspiration from denoising methods designed for additive noise, various heuristic despeckling approaches—such as non-local means \cite{deledalle2009iterative}, SAR-BM3D \cite{parrilli2011nonlocal}, and neural network-based solutions \cite{zhu2021deep}—have been widely adopted, particularly for high-resolution imaging tasks.
For instance, the DnCNN  architecture \cite{zhang2017beyond} has been trained for despeckling images, both with the log-transformation \cite{chierchia2017sarcnn} and without it \cite{wang2017idcnn}. Also, self-supervised despeckling algorithms, based on the Noise2Noise \cite{lehtinen2018noise2noise} and Noise2Void \cite{krull2019noise2void} frameworks originally developed for additive denoising, are extended to the despeckling problem in \cite{dalsasso2021sar2sar} and \cite{molini2021speckle2void}, respectively.
State-of-the-art generative models have also been explored for addressing the image despeckling problem. These include the use of generative adversarial networks as demonstrated in \cite{bobrow2019deeplsr} and diffusion-based probabilistic models as explored in \cite{guha2023sddpm}.

\subsection{Our contributions}

We propose Bayesian Despeckling via QMAP (BD-QMAP), a novel despeckling algorithm inspired by the Quantized Maximum a Posteriori (Q-MAP) estimator \cite{zhou2023bayesian}, originally developed for additive noise. QMAP defines the estimator as the minimizer of a cost function over the space of all possible solutions within the signal's support space $\Xc^n$. The cost function consists of two terms: a negative log-likelihood term capturing the noise model and a regularization term enforcing distributional similarity between the quantized reconstruction and quantized ground-truth data. By leveraging learned statistics, QMAP assigns weights to unique realizations of the reconstruction, effectively reducing the impact of irrelevant solutions. 

Building on these ideas, BD-QMAP adapts QMAP principles to multiplicative noise, tailoring the framework for general structured sources. To clarify its operation, we simplify the formulation for classic structured sources, including memoryless sources with a mixture of continuous and discrete components and piecewise-constant sources. We establish a theoretical lower bound on BD-QMAP’s performance for piecewise-constant first-order Markov processes. Experimental results demonstrate that BD-QMAP achieves state-of-the-art performance on piecewise-constant sources modeled as first-order Markov processes.

\subsection{Notations and definitions}

Finite sets are denoted by calligraphic letters. For a finite set $\Ac$, $|\Ac|$ denotes its size. For $b\in\mathbb{N}^+$, the $b$-bit quantized version of $x\in\mathbb{R}$ is denoted as $[x]_b$,  defined as $[x]_b=2^{-b}\lfloor 2^b x\rfloor$. For $x^k\in\Xc^k$, $[x^k]_b$ denotes the element-wise $b$-bit quantization of $x^k$. The $b$-bit quantized version of $\Xc\subset \mathbb{R}$ is denoted by $\Xc_b$, defined as 
\[
\Xc_b=\{[x]_b:\;x\in\Xc\}.
\]

Consider sequence $u^n\in\Uc^n$, with finite alphabet $\mathcal{U}$ ($|\Uc|<\infty$). The $k$-th order
empirical distribution of $u^n$,  $\hat{p}^{k}(\cdot|u^n)$,  is defined as follows. For  $a^{k}\in\mathcal{U}^{k}$,
\begin{align}
\hat{p}^{k}(a^{k}|u^n)  = \frac{1}{n-k+1} \sum_{i=1}^{n-k+1} \mathds{1}_{u_i^{i+k-1}=a^{k}}.\label{eq:emp-dist}
\end{align}


\subsection{Paper organization}

This paper is organized as follows: \Cref{sec:bd-qmap} introduces the BD-QMAP despeckler. \Cref{sec:classical-structured} discusses its application to two classical structured sources and derives a lower bound for the MSE of the piecewise constant source. \Cref{sec:experiments} describes numerical experiments, including the algorithm's implementation and a performance comparison with other despeckling methods.

\section{Bayesian Despeckling via QMAP}\label{sec:bd-qmap}

Consider the despeckling problem, where the goal is to recover $X^n$ from noisy measurements $Y^n=X^nW^n$. Here,  $X^n$ follows a known prior distribution $p(x^n)$ and $W^n$ is i.i.d.~$\Nc(0,\sigma_w^2)$. In this formulation, we neglect the effect of additive noise, focusing solely on the multiplicative noise model, as is commonly done in coherent imaging applications. Note that under this model, the variance of speckle noise does not play a role, as one can always scale the measurements by $1/\sigma_w$. Therefore, in the remainder of the paper, without loss of generality, we assume that $\sigma_w^2=1$.

Inspired by the Q-MAP denoiser, introduced for additive noise \cite{zhou2023bayesian},  we propose BD-QMAP, a novel despeckling algorithm that is applicable to general structured sources. To define this approach, first we review how the structure of the source is accounted for in the original Q-MAP algorithm.  For  $a^k\in\Uc_b^k$, define $w_{a^k}>0$ as 
\[
w_{a^k}=-\log \P([X^k]_b=a^k),
\]
where the probability $P$ is computed with respect to the known distribution $p(x^n)$. 
Here, $k\in\mathbb{N}^+$ and $b\in\mathbb{N}^+$  denote the memory parameter and the quantization level, respectively. Then, given weights $\mathbf{w}=(w_{a^k}:a^k\in\Uc_b^k)$, the weight assigned to sequence $u^n\in\Uc^n$ is defined as 
\begin{equation}\label{eq:qmap-weights}
  c_{\mathbf{w}}(u^n)=\sum_{a^k\in\mathcal{X}^k_b} w_{a^k}\hat{p}^{k}(a^{k}|u^n).  
\end{equation}
Finally, the BD-QMAP method, recovers $X^n$ as $\hat{X}^{n, (k,b)}$  defined as 
\begin{equation} \label{eq:qmap-despeckle}
   \hat{X}^{n, (k,b)}= \argmin_{u^n\in\mathcal{X}^n} {1\over n}\sum_{i=1}^n \big(\log u_i^2+\frac{Y_i^2}{u_i^2}\big)+\frac{\lambda}{b}c_{\mathbf{w}}(u^n).
\end{equation}
The cost function in \eqref{eq:qmap-despeckle} comprises of two terms. The first term represents a fidelity criterion derived from the negative  likelihood of the observations given the signal under the multiplicative noise model. Note that given $X^n=x^n$, $Y^n\sim\Nc({\bf 0}, X^2)$, where $X=\diag(x^n)$. Therefore, 
\[
-\log p(y^n|x^n)={1\over 2}\sum_{i=1}^n(\log x_i^2+{y_i^2\over  x_i^2})+C.
\]
where $C$ is a constant not depending on $x^n$ or $y^n$.
The second term in \eqref{eq:qmap-despeckle} imposes a prior on the ground-truth signal using the set of weights assigned to the quantized representation of the candidate reconstruction. The weights defined in \eqref{eq:qmap-weights} are a function of the signal's known distribution. In other words, the weights are expected to summarize the source's $n$-dimensional distribution characterized by $p(x^n)$ into a finite number $|\Xc_b|^k$ of  positive weights. 

A key advantage of this approach is that it leads to a tractable approach for modeling and utilizing the source's structure. It can be shown that for structured sources, i.e., the sources with information dimension strictly less than one \cite{wu2010renyi,jalali2017universal}, identified by having singularities in their distributions, e.g., spike and slap distribution, the weights $\wv$ can be divided into two groups: 1) A small set of weights that identify the key structure of the source, and 2) a large set of weights that have negligible impact in the optimization \cite{zhou2023bayesian}.

\begin{remark}\label{rmk:opt-challenge}
Solving the BD-QMAP optimization is challenging due to the nature of the cost function, which combines a non-convex loss term derived from the log-likelihood and a regularization term defined on the discretized space of sequences. To address this complexity, the optimization can be simplified by restricting the search space to $\Xc_b^n$, the space of quantized sequences. As will be explained in Section~\ref{subsec:alg-considerations}, this restriction enables the application of the Viterbi algorithm for efficient optimization. While this approach is suboptimal compared to solving the original problem, it offers significant computational advantages. Moreover, for cases such as the piecewise-constant source studied later, this simplification facilitates learning the structure of the jumps, allowing the original optimization to be solved in the continuous space with improved computational efficiency.
\end{remark}

To gain deeper insight into the BD-QMAP optimization and the roles of its two terms, we examine two classic structured processes in the next section: structured i.i.d.~sources and piecewise-constant sources. For each source, we derive a simplified form of the BD-QMAP despeckler optimization. Moreover, for the piecewise-constant source, we theoretically establish a lower bound on the expected mean squared error (MSE) achievable by the BD-QMAP method.


\section{Analysis of BD-QMAP for classic structured sources}\label{sec:classical-structured}

In the following, we focus on two classic structured source models and study BD-QMAP under each model. 

\subsection{Structured memoryless source}

Consider an i.i.d. process ${\bf X}$ such that $X_i\sim (1-q_0)\delta_{x_m}+q_0{\rm Uniform}(x_{m},x_M)$, Setting $k=1$ and using the results of \cite[Sec. 3.1]{zhou2023bayesian} to simplify the second term in \eqref{eq:qmap-despeckle}, the BD-QMAP algorithm  can be written as 
 \[
   { \hat{X}^{n, (1, b)} = \arg\min_{u^n\in\Uc^n}  \Big[{1\over n}\sum_{i=1}^n \big(\log u_i^2+\frac{Y_i^2}{u_i^2}\big)+ \lambda (1+\gamma)(1- \hat{p}^{1}(x_m|[u^n]_b))  \Big]},
\]
  where $\gamma={1\over b} \log({1-q_0\over q_0}+ 2^{-b})=O({1\over b})$ . 
Note that $1- \hat{p}^{1}(x_m|[u^n]_b) ={1\over n}\sum_{i=1}^n\mathds{1}_{[u_i]_b\neq x_{m}}$. Therefore, the optimization simplifies to a symbol-by-symbol optimization as follows: 
 \[
   { \hat{X}_i^{ (1, b)} = \arg\min_{u\in\Uc}  \Big[\log u^2+\frac{Y_i^2}{u^2}+  \lambda (1+\gamma)\mathds{1}_{[u]_b\neq x_m}   \Big]}.
\]
Note that if  $[u]_b\neq x_m$, the loss function is minimized at $\xh_i=|Y_i|$. Therefore, assuming that $b$ is large, to solve the optimization one needs to compare $\log Y_i^2+1+\lambda \gtrless \log x_m^2+{Y_i^2\over x_m^2}$, or 
\begin{align}
1+\lambda \gtrless ({Y_i\over x_m})^2-\log ({Y_i\over x_m})^2.\label{eq:qmap-I}
\end{align}
This optimization has the following closed-form solution
\[
\hat{X}_i^{ (1, b)}= |Y_i|\mathds{1}_{|Y_i|/x_m\notin \Ic_{\lambda}}+x_m\mathds{1}_{|Y_i|/x_m\in \Ic_{\lambda}},
\]
where $\Ic_{\lambda}$ denotes an interval around one that is defined by \eqref{eq:qmap-I}.

\subsection{Piecewise constant source}\label{sec:piecewise}
Consider a  first order Markov process ${\bf X}$  such that given $X_{i}=x_i$, $X_{i+1}$ is distributed  as $(1-q_0)\delta_{x_i}+q_0{\rm Uniform}(x_{m},x_M)$. Setting $k=2$ and again using the results of \cite[Sec. 3.2]{zhou2023bayesian}, the BD-QMAP  algorithm  simplifies to  
\begin{align}
     \hat{X}^{n, (2, b)}=\arg\min_{u^n\in\mathcal{X}^n}\Big[{1\over n}\sum_{i=1}^n \big(\log u_i^2+\frac{Y_i^2}{u_i^2}\big)+(\lambda+{1\over b}\eta){N_{ J}([u^n]_b)\over n-1}\Big],\label{eq:QMAP-Markov-main}
  \end{align}
  where $\eta=-\lambda \log q_0-  \log(1-q_0+q_02^{-b})$ is a constant not depending on $u^n$. Here, $N_{ J}(u^n)=|\{i: u_i\neq u_{i+1}\}|$ denotes the number of jumps in $u^n$.

To provide a clearer understanding of how the BD-QMAP loss function operates, as well as the roles of its two key terms—the first ensuring reconstruction fidelity to the observations, and the second acting as a regularizer to promote the source structure—the following lemma offers an alternative representation of \eqref{eq:QMAP-Markov-main}. 

Consider $\nv=(n_1,\ldots,n_{k+1})\in(\mathbb{N}^+)^{k+1}$, such that  $\sum_{j=1}^{k+1}n_j=n$. Then,  for any $j\in\{1,\dots,k+1\}$, define
\begin{align}
\Ic_j(\nv)=\Big\{\sum_{i=1}^{j-1}n_i + 1,\ldots, \sum_{i=1}^{j}n_i\Big\}.\label{eq:def-Ik}
\end{align}

\begin{lemma}\label{lemma:Markov-QMAP-opt}
Solving the optimization in \eqref{eq:QMAP-Markov-main} is equivalent  to solving the following optimization 
\begin{align}
     \hat{X}^{n, (2, b)}= \min_{k\in\{0,\ldots,n-1\}}\Big(\min_{(n_1,\ldots,n_{k+1})} \sum_{j=1}^{k+1} n_j\log \big({1\over n_j}\sum_{l\in\Ic_j}Y_l^2\big)+{n\over n-1}(\lambda+{1\over b}) k\Big),\label{eq:QMAP-v2-Markov}
 \end{align}
 where for any $k\in\{0,\ldots,n-1\}$, the inner optimization is over $(n_1,\ldots,n_{k+1})\in(\mathbb{N}^+)^{k+1}$, such that  $\sum_{i=1}^{k+1}n_i=n$.  Moreover, for $i\in \Ic_j(\nv)$, 
\begin{align}\label{eq:constant-estimator}
\hat{X}^{(2, b)}_i= \sqrt{{1\over n_j}\sum_{l\in\Ic_j}Y_l^2}.
\end{align}
\end{lemma}
In other words, Lemma \ref{lemma:Markov-QMAP-opt} implies that solving the BD-QMAP algorithm is equivalent to identifying the locations of the jumps such that, over each constant interval, the source input is estimated through appropriate averaging of the noisy observations, thereby minimizing the corresponding loss function. It is important to emphasize that if the regularization term enforcing the source structure were absent (i.e., if $\lambda=0$), the concavity of the $\log$ function would result in the solution equaling the absolute value of input noisy sequence, with the maximal number of jumps. On the other hand, choosing $\lambda$ too large, will ensure that the output has no jumps.

The following theorem characterizes a  lower bound on the performance achievable by BD-QMAP optimization \eqref{eq:QMAP-Markov-main}.

\begin{theorem}\label{thm:MSE-lower-bound}
    Consider $X^n$ generated by a stationary first-order Markov source $\mathbf{X}=\{X_i\}_{i\geq 1}$, characterized by  $p(X_{i+1}=x_{i+1}|X_{i}=x_{i})=(1-q_0)\delta_{x_i}+q_0{\rm \pi_c}(x_{m},x_M)$, where $\pi_c$ denotes the pdf of an absolutely continuous distribution with bounded support  $[x_{m}, x_M]$, $x_m>0$.  Let $\Xh^n$ denote the solution of BD-QMAP optimization, when the number of jumps are  known to  be $k=k(n)$.   Then,
\begin{align}\label{eq:thm1-result}
\E[{1\over n}\|X^n-\Xh^{n}\|_2^2] 
\geq 2q_0\eta^2 \Big(q_0c_1+q_0^2c_2
+\E[T(1- {1\over \sqrt{\ex}} {(1-{1\over T})^{T\over 2} \over (1-{2\over T})^{T-1 \over 2}})\mathds{1}_{T\geq 3}]\Big)-\upsilon_n,
\end{align}
where $\eta^2=\E[X_i^2]$, $c_1=3-\sqrt{2/\pi}-2\sqrt{\pi}$ and $c_2=\sqrt{\pi}-2$, and $T\sim\operatorname{Geometric}(q_0)$. Here, 
\begin{align*}
\upsilon_n=2x_{M}^2
\Big(
 {\epsilon \over 1-\epsilon}
 +
nq_0 \ex^{-\lfloor nq_0(1+\epsilon)\rfloor\left(q_0t_1+\ln \frac{1}{1+q_0t_1}\right)}
 +
 nq_0 \ex^{-\lfloor nq_0(1-\epsilon)\rfloor\left(q_0t_2+\ln \frac{1}{1+q_0t_2}\right)}
\Big),
\end{align*}
with $t_1={1 /q_0}-{n/ \lfloor nq_0(1+\epsilon)\rfloor}$ and $t_2={n/\lfloor nq_0(1-\epsilon)\rfloor}-{1/q_0}$.
\end{theorem}

The following corollary, presents a simplified form of Theorem \ref{thm:MSE-lower-bound}, for large values of $n$.
\begin{corollary}\label{cor:mse-bound-convergence}
Consider the same setup as in Theorem \ref{thm:MSE-lower-bound}. Then, 
\begin{align*}
\E[{1\over n}\|X^n-\Xh^{n}\|_2^2] 
\geq &2q_0\eta^2 (q_0c_1-q_0^2c_2
+\E[T(1- {1\over \sqrt{\ex}} {(1-{1\over T})^{T\over 2} \over (1-{2\over T})^{T-1 \over 2}})\mathds{1}_{T\geq 3}])-\upsilon_n,
\end{align*}
where $\upsilon_n=O(n^{-1/4})$.
\end{corollary}

Finally, to better understand the term in \eqref{eq:thm1-result} that involves  expectation over $T$, note that for small values of $q_0$, with high probability,   $T$ is large. On the other hand, for large values of $T$, $f(T)=1/4+O(T^{-1})$, as
 \begin{align*}
     f(T)=T(1- {1\over \sqrt{\ex}} {(1-{1\over T})^{T\over 2} \over (1-{2\over T})^{T-1 \over 2}})
     &=T(1-\exp(-\frac{1}{2}+\frac{T}{2}\ln(1-\frac{1}{T})-\frac{T-1}{2}\ln(1-\frac{2}{T})))\\
     &=T(1-\exp(-\frac{1}{2}+\frac{T}{2}(-\frac{1}{T}-\frac{1}{2T^2}+O(T^{-3}))-\frac{T-1}{2}(-\frac{2}{T}-\frac{2}{T^2}+O(T^{-3}))))\\
     &=T(1-\exp(-\frac{1}{4T}-\frac{1}{T^2}+O(T^{-2})))\\
     &=T(1-1+\frac{1}{4T}+O(T^{-2}))\\
     &=\frac{1}{4}+O(T^{-1}).
 \end{align*}
\section{Proofs} \label{sec:proofs}
The following lemma will be used in the proof of the \Cref{thm:MSE-lower-bound}.
\begin{lemma}[Concentration of Geometric \cite{janson2018tail}]\label{lemma:tail-sum-geometric}
Assume that $T_i\overset{iid}{\sim}\operatorname{Geometric}(q_0)$, $i=1,\ldots,N$.   Then for any $t>0$
\begin{align*}
    \P\left(\left|\frac{\sum_{i=1}^NT_i}{N}-\frac{1}{q_0}\right|>t\right)\leq2e^{-N\left(q_0t+\ln \frac{1}{1+q_0t}\right)}.
\end{align*}
\end{lemma}

\subsection{Proof of Lemma \ref{lemma:Markov-QMAP-opt}}
Note that the set $\Uc^n$ can be partitioned into $n$ subsets depending on the number of jumps. That is, $\Uc^n=\bigcup_{k=0}^{n-1}\Uc_{k},$
where, for $k=0,\ldots,n-1$, 
\[
\Uc_k=\{u^n\in\Uc^n:\;\sum_{i=2}^n \mathds{1}_{u_i\neq u_{i-1}}=k\}.
\]
Then, the optimization in \eqref{eq:QMAP-Markov-main} can also be solved as follows
\begin{align}
&\min_{k\in\{0,\ldots,n-1\}}\min_{u^n\in \Uc_k}[{1\over n}\sum_{j=1}^n(\log u_i^2+{Y_i^2\over u_i^2})+{1\over n-1}(\lambda+{1\over b}\eta) \sum_{i=2}^n\mathds{1}_{u_i\neq u_{i-1}}]\nonumber\\
&\equiv \min_{k\in\{0,\ldots,n-1\}}\Big(\min_{u^n\in \Uc_k}\sum_{j=1}^n\big(\log u_i^2+{Y_i^2\over u_i^2}\big)+{n\over n-1}(\lambda+{1\over b}) k\Big).\label{eq:QMAP-Uk-union}
\end{align}
First, consider the inner optimization which is over all sequences with exactly $k$ jumps:
\[
\min_{u^n\in \Uc_k}\big(\log u_i^2+{Y_i^2\over u_i^2}\big).
\]
Consider a sequence with constant intervals of  length $n_1,\ldots,n_{k+1}$ with corresponding values $a_1,\ldots,a_{k+1}$. (Clearly $\sum_{i=1}^{k+1}n_i=n$.) That is, $u^n$ with $k$ jumps is written as 
\[
u^n=\underbrace{a_1, \ldots, a_1}_{n_1},\underbrace{a_2, \ldots, a_2}_{n_2},\ldots, \underbrace{a_{k+1}, \ldots, a_{k+1}}_{n_{k+1}}.
\]
Then, the optimization can be written as 
\begin{align}
\min_{(n_1,\ldots,n_{k+1}), (a_1,\ldots,a_{k+1})\in\Uc^{k+1}} \big(\sum_{j=1}^{k+1} (n_j\log a_i^2+{1\over a_j^2}\sum_{l\in\Ic_j}Y_l^2)\big).\label{eq:QMAP-Uk}
\end{align}
where $\Ic_j$ is defined as \eqref{eq:def-Ik}. Then, it is easy to see that fixing the intervals, the optimal values can be found as
\[
\hat{a}_j^2={1\over n_j}\sum_{l\in\Ic_j}Y_l^2.
\]
Using this observation, the optimization in \eqref{eq:QMAP-Uk} can be written as 
\begin{align}
\min_{(n_1,\ldots,n_{k+1})} \sum_{j=1}^{k+1} n_j\log \big({1\over n_j}\sum_{l\in\Ic_j}Y_l^2\big),
\end{align}
which yields the desired result.

\subsection{Proof of Theorem \ref{thm:MSE-lower-bound}}
Assuming that $X^n$ contains $k=k(n)$ jumps, it can be written as  
\[
X^n=\underbrace{\alpha_1, \ldots, \alpha_1}_{T_1},\underbrace{\alpha_2, \ldots, \alpha_2}_{T_2},\ldots, \underbrace{\alpha_{k+1}, \ldots, \alpha_{k+1}}_{T_{k+1}}.
\]
Moreover, when the number of jumps ($k$) is known apriori by the algorithm, \eqref{eq:QMAP-v2-Markov} simplifies to  
\begin{align}
\Xh^n= \min_{(n_1,\ldots,n_{k+1}):\;\sum_{i=1}^{k+1}n_i=n} \sum_{j=1}^{k+1} n_j\log \big({1\over n_j}\sum_{l\in\Ic_j}Y_l^2\big).\label{eq:thm2-1}
  \end{align}
Clearly, the expected error is minimized when the locations of the jumps are detected correctly (i.e., $n_i=T_i$, $i=1,\ldots,n+1$). We refer to this solution as the maximum likelihood (ML) solution, as it coincides with the ML solution when the number of jumps and their locations are known.   As we just argued,
\[
\E[{1\over n}\|X^n-\Xh^n\|_2^2]\geq \E[{1\over n}\|X^n-\Xh^{n,{\rm (ML)}}\|_2^2].
\]  
For $j=1,\ldots,k(n)+1$, define 
\begin{align}
\hat{\alpha}_j=\sqrt{{1\over T_j}\sum_{l\in\Ic_j}Y_l^2}=\alpha_j\sqrt{{1\over T_j}\sum_{l\in\Ic_j}W_l^2}.
\end{align} 
Then, for $i\in\Ic_j$,
  \begin{align}
  \hat{X}_i^{\rm (ML)}=\hat{\alpha}_j,
  \end{align}
  which implies that 
\begin{align}
{1\over n}\|X^n-\Xh^{n,{\rm (ML)}}\|_2^2
&={1\over n}\sum_{j=1}^k\alpha_j^2T_j\left(1-\sqrt{{1\over T_j}\sum_{l\in\Ic_j}W_l^2}\right)^2+T_{k+1}{\alpha_{k+1}^2\over n}\left(1-\sqrt{{1\over T_{k+1}}\sum_{l\in\Ic_{k+1}}W_l^2}\right)^2\nonumber\\
&\geq {1\over n}\sum_{j=1}^k\alpha_j^2T_j\left(1-\sqrt{{1\over T_j}\sum_{l\in\Ic_j}W_l^2}\right)^2\nonumber\\
&= {q_0\over k}\sum_{j=1}^k\alpha_j^2T_j\left(1-\sqrt{{1\over T_j}\sum_{l\in\Ic_j}W_l^2}\right)^2+({k\over n}-q_0){1\over k}\sum_{j=1}^k\alpha_j^2T_j\left(1-\sqrt{{1\over T_j}\sum_{l\in\Ic_j}W_l^2}\right)^2\nonumber\\
&\stackrel{\rm (a)}{\geq}  {q_0\over k}\sum_{j=1}^k\alpha_j^2T_j\left(1-\sqrt{{1\over T_j}\sum_{l\in\Ic_j}W_l^2}\right)^2-|{k\over n}-q_0|{x_M^2\over k}\sum_{j=1}^kT_j(1+{1\over T_j}\sum_{l\in\Ic_j}W_l^2)\nonumber\\
&\stackrel{\rm (b)}{\geq}  {q_0\over k}\sum_{j=1}^k\alpha_j^2T_j\left(1-\sqrt{{1\over T_j}\sum_{l\in\Ic_j}W_l^2}\right)^2-|{1\over q_0}-{n\over k}|x_M^2q_0(1+{1\over n}\sum_{l=1}^n W_l^2),
\label{eq:thm1-main-lower-bound}
\end{align}  
where (a) follows because $(a-b)^2\leq a^2+b^2$ for all $a,b$ and (b) holds because $\sum_{j=1}^kT_j\leq n$ and $\sum_{j=1}^k\sum_{l\in\Ic_j}W_l^2\leq \sum_{l=1}^nW_l^2$.

Taking the expected value of both sides of \eqref{eq:thm1-main-lower-bound}, it follows that 
\begin{align}
\E[{1\over n}\|X^n-\Xh^{n,{\rm (ML)}}\|_2^2]
& \geq q_0\eta^2 \E\left[T\Big(1-\sqrt{{1\over T}\sum_{i=1}^TW_i^2}\;\Big)^2 \right]-x_M^2q_0\E\left[|{1\over q_0}-{n\over k}|(1+{1\over n}\sum_{l=1}^n W_l^2)\right]\nonumber\\
& = q_0\eta^2 \E\left[T\Big(1-\sqrt{{1\over T}\sum_{i=1}^TW_i^2}\;\Big)^2 \right]-2x_M^2q_0\E\left[|{1\over q_0}-{n\over k}|\right],\label{eq:thm2-main-bound}
\end{align}
where the last line follows because the source $\mathbf{X}$ and the speckle noise process are independent. 
We first focus on the first term. Note that 
\begin{align}
\E\left[T\Big(1-\sqrt{{1\over T}\sum_{j=1}^{T}w_j^2}\;\Big)^2\right]
&=\E\left[T\Big(1+{1\over T}\sum_{j=1}^{T}w_j^2-2\sqrt{{1\over T_i}\sum_{j=1}^{T_i}w_j^2}\Big)\right]\nonumber\\
&=\E\left[2T-2T\sqrt{{1\over T}\sum_{j=1}^{T}w_j^2}\;\right]\nonumber\\
&=2\E\left[T-\sqrt{2T}{\Gamma({T+1\over 2})\over \Gamma({T\over 2})}\right],\label{eq:thm2-2}
\end{align}    
where the last line follows because
\begin{align}
\E\left[\sqrt{T\sum_{j=1}^{T}w_j^2}\;\right]
&=\E\left[\E\left[\sqrt{T\sum_{j=1}^{T}w_j^2}\;\bigg|\;T\right]\right]= \E\left[\sqrt{2T}{\Gamma({T+1\over 2})\over \Gamma({T\over 2})}\right].
\end{align}

 It can be shown that, for all $x>0$,
\begin{align}
\sqrt{2\pi} {x}^{x+0.5}\ex^{-x}\ex^{{1\over 1+12x}} \leq \Gamma(x+1)\leq \sqrt{2\pi} {x}^{x+0.5}\ex^{-x}\ex^{{1\over 12x}}.
\end{align}
Therefore, for $T>2$, 
\begin{align}
{\Gamma({T+1\over 2})\over \Gamma({T\over 2})} 
\;\leq\;&  
{ {({T-1\over 2}})^{T\over 2}\ex^{-(T-1)/2}\ex^{{1\over 6(T-1)}}\over {({T\over 2}-1})^{T-1 \over 2}\ex^{-T/2+1}\ex^{{1\over 6T-11}}}\nonumber\\
\;=\;&\ex^{-{1\over 2}}\sqrt{T\over 2}(1-{1\over T})^{T\over 2}(1-{2\over T})^{-{T-1 \over 2}}\ex^{{1\over 6(T-1)}-{1\over 6T-11}}\nonumber\\
\;=\;&
\ex^{-{1\over 2}}\sqrt{T\over 2} {\rm exp}\Big({T\over 2}\ln(1-{1\over T})-{T-1 \over 2}\ln(1-{2\over T})+{1\over 6(T-1)}-{1\over 6T-11}\Big)\nonumber\\
\;\leq \;&
\ex^{-{1\over 2}}\sqrt{T\over 2} {\rm exp}\Big({T\over 2}\ln(1-{1\over T})-{T-1 \over 2}\ln(1-{2\over T})\Big)\nonumber\\
\;= \;& \sqrt{T\over 2\ex} {(1-{1\over T})^{T\over 2} \over (1-{2\over T})^{T-1 \over 2}}.
\end{align}
Therefore, we can bound \eqref{eq:thm2-2} as
\begin{align}\label{eq:thm2-mse-first-term}
2\E\left[T-\sqrt{2T}{\Gamma({T+1\over 2})\over \Gamma({T\over 2})}\right]
&\geq  
2q_0(1-{\sqrt{2}\over \Gamma(0.5)})
+2q_0(1-q_0)(2-2\Gamma(1.5))
+ 2\E[T(1- {1\over \sqrt{\ex}} {(1-{1\over T})^{T\over 2} \over (1-{2\over T})^{T-1 \over 2}})\mathds{1}_{T\geq 3}]
\end{align}

Combining \eqref{eq:thm2-main-bound} and   \eqref{eq:thm2-mse-first-term}, it follows that 
\begin{align}
\E[{1\over n}\|X^n-\Xh^{n,{\rm (ML)}}\|_2^2] 
\geq q_0&\eta^2 (2q_0(1-{\sqrt{2}\over \Gamma(0.5)})
+2q_0(1-q_0)(2-2\Gamma(1.5))\\
&+2\E[T(1- {1\over \sqrt{\ex}} {(1-{1\over T})^{T\over 2} \over (1-{2\over T})^{T-1 \over 2}})\mathds{1}_{T\geq 3}])
- 2x_M^2q_0\E\left[|{1\over q_0}-{n\over k}|\right].\label{eq:thm2-main-bound-v2}
\end{align}




To bound the last term, note that $k$ is a random variable that depends on $n$. Given $\epsilon>0$, define event $\Ec$ as
\begin{align}
\Ec=\{nq_0(1-\epsilon)\leq k(n)\leq nq_0(1+\epsilon)\}.
\end{align}
Note that conditioned on $\Ec$, we have $1-\epsilon \leq { k(n)\over nq_0} \leq 1+\epsilon$, which implies that 
\[
{1\over q_0(1+\epsilon)}-{1\over q_0} \leq {n\over k} -{1\over q_0}\leq {1\over q_0(1-\epsilon)}-{1\over q_0}.
\]
Using this observation, we can bound the remaining  term in \eqref{eq:thm2-main-bound-v2}, as follows:
\begin{align}
\E\left[|{1\over q_0}-{n\over k}|\right]
&=\E\left[|{1\over q_0}-{n\over k}|(\mathds{1}_{\Ec}+\mathds{1}_{\Ec^c})\right]\nonumber\\
&\leq {\epsilon \over q_o(1-\epsilon)}+nP(\Ec^c).\label{eq:error-bound}
\end{align}

Finally, to bound $P(\Ec^c)$, note that 
\begin{align}
\P(k(n)\geq nq_0(1+\epsilon))
&= \P(\sum_{i=1}^{\lfloor nq_0(1+\epsilon)\rfloor} T_i<n)\nonumber\\
&= \P\left({1\over \lfloor nq_0(1+\epsilon)\rfloor}\sum_{i=1}^{\lfloor nq_0(1+\epsilon)\rfloor} T_i-{1\over q_0}<{n\over \lfloor nq_0(1+\epsilon)\rfloor}-{1\over q_0}\right)\nonumber\\
&\leq \P\left(|{1\over q_0}-{1\over \lfloor nq_0(1+\epsilon)\rfloor}\sum_{i=1}^{\lfloor nq_0(1+\epsilon)\rfloor} T_i|>{1\over q_0}-{n\over \lfloor nq_0(1+\epsilon)\rfloor}\right)\nonumber\\
&\leq \ex^{-\lfloor nq_0(1+\epsilon)\rfloor\left(q_0t_1+\ln \frac{1}{1+q_0t_1}\right)},\label{eq:P-Ec-1}
\end{align}
where the last line follows from Lemma \ref{lemma:tail-sum-geometric}, with
\[
t_1={1\over q_0}-{n\over \lfloor nq_0(1+\epsilon)\rfloor}.
\]
Similarly, again using Lemma \ref{lemma:tail-sum-geometric}, we have 
\begin{align}
\P(k(n)\leq nq_0(1-\epsilon))
&=\P(\sum_{i=1}^{\lfloor nq_0(1-\epsilon)\rfloor} T_i>n)\nonumber\\
&=\P({1\over \lfloor nq_0(1-\epsilon)\rfloor} \sum_{i=1}^{\lfloor nq_0(1-\epsilon)\rfloor} T_i-{1\over q_0}>{n\over \lfloor nq_0(1-\epsilon)\rfloor}-{1\over q_0})\nonumber\\
&\leq \ex^{-\lfloor nq_0(1-\epsilon)\rfloor\left(q_0t_2+\ln \frac{1}{1+q_0t_2}\right)},\label{eq:P-Ec-2}
\end{align}
with
\[
t_2={n\over \lfloor nq_0(1-\epsilon)\rfloor}-{1\over q_0}.
\]
Combining \eqref{eq:error-bound}, \eqref{eq:P-Ec-1} and \eqref{eq:P-Ec-2} with \eqref{eq:thm2-main-bound-v2} yields the desired result. That is, 
\begin{align}
\E[{1\over n}\|X^n-\Xh^{n}\|_2^2] 
\geq &q_0\eta^2 (2q_0(1-{\sqrt{2}\over \Gamma(0.5)})
+2q_0(1-q_0)(2-2\Gamma(1.5))
+2\E[T(1- {1\over \sqrt{\ex}} {(1-{1\over T})^{T\over 2} \over (1-{2\over T})^{T-1 \over 2}})\mathds{1}_{T\geq 3}])\\ &- 2x_M^2q_0
\Big(
 {\epsilon \over q_o(1-\epsilon)}
 +
n \ex^{-\lfloor nq_0(1+\epsilon)\rfloor\left(q_0t_1+\ln \frac{1}{1+q_0t_1}\right)}
 +
 n \ex^{-\lfloor nq_0(1-\epsilon)\rfloor\left(q_0t_2+\ln \frac{1}{1+q_0t_2}\right)}
\Big).
\end{align}

\subsection{Proof of \Cref{cor:mse-bound-convergence}}
For $x\geq 0$ we can show \cite{topsoe2007some} that $\ln(1+x)\leq x\frac{x+6}{2x+6}$ thus,
\begin{align}\label{eq:cor-log-bound}
    x-\ln(1+x)\geq\frac{x^2}{2x+6}.
\end{align}

In addition, for $x\geq0$ we have
\begin{align}\label{eq:cor-omega1}
    1-\frac{x}{\lfloor x(1+x^{-1/4})\rfloor}=\Omega(\frac{1}{x^{1/3}}),
\end{align}
and,
\begin{align}\label{eq:cor-omega2}
    \frac{x}{\lfloor x(1-x^{-1/4})\rfloor}-1=\Omega(\frac{1}{x^{1/3}}).
\end{align}

\eqref{eq:cor-omega1} holds for $x\geq x_0$,
\begin{align*}
    1-\frac{x}{\lfloor x+x^{3/4}\rfloor}
    &=1-\frac{x}{\lfloor x+x^{3/4}\rfloor-x-x^{3/4}+x+x^{3/4}}\\
    &\geq1-\frac{x}{-1+x+x^{3/4}}\\
    &=\frac{x^{3/4}-1}{x+x^{3/4}-1}\\
    &=\frac{1}{x^{1/3}}\underbrace{\left(\frac{x^{3/4}-1}{x^{2/3}+x^{5/12}-x^{-1/3}}\right)}_{(*)}\\
    &\geq \frac{1}{x^{1/3}},
\end{align*}
where $x_0$ is some point after which $(*)$ remains greater than one. Similar steps proves \eqref{eq:cor-omega2}. 

Let $\epsilon=(nq_0)^{-1/4}$ in the result of \Cref{thm:MSE-lower-bound}, using \eqref{eq:cor-omega1}, \eqref{eq:cor-omega2} we have $q_0t_1=\Omega(n^{-1/3})$ and $q_0t_2=\Omega(n^{-1/3})$, and applying \eqref{eq:cor-log-bound} we have
\begin{align*}
\E[{1\over n}\|X^n-\Xh^{n}\|_2^2] 
\geq &2q_0\eta^2 \Big(q_0c_1+q_0^2c_2
+\E[T(1- {1\over \sqrt{\ex}} {(1-{1\over T})^{T\over 2} \over (1-{2\over T})^{T-1 \over 2}})\mathds{1}_{T\geq 3}]\Big)\\ &- 2x_M^2
\Big(
 {(nq_0)^{-1/4} \over 1-(nq_0)^{-1/4}}
 +
q_0n \ex^{-\lfloor nq_0(1+(nq_0)^{-1/4})\rfloor\left(d_1^2n^{-2/3}\over 2n^{-1/3} +6\right)}
 +
 q_0n \ex^{-\lfloor nq_0(1-(nq_0)^{-1/4})\rfloor\left(d_2^2n^{-2/3}\over 2n^{-1/3} +6\right)}
\Big),
\end{align*}
where $d_1,d_2\in\mathds{R}^+$ are constants. Second line converges to zero as $n\rightarrow\infty$ with $O(n^{-1/4})$.

\section{Numerical Experiments}\label{sec:experiments}

This section evaluates the performance of the proposed BD-QMAP method. We begin by addressing the practical implementation challenges of the BD-QMAP optimization problem. Next, we compare its performance against the theoretical lower bound derived in \Cref{thm:MSE-lower-bound}. Finally, we investigate BD-QMAP’s application to piecewise-constant sources.

\subsection{Implementation of BD-QMAP}\label{subsec:alg-considerations}

As noted in \Cref{rmk:opt-challenge}, the practical implementation of BD-QMAP must address the distinct search spaces of the log-likelihood term and the regularizer. Restricting the search space to the quantized domain ensures computational feasibility, enabling discrete suboptimal optimization. Specifically, we limit the search space in $ \eqref{eq:qmap-despeckle} $ to $b$-bit quantized sequences, denoted by $\mathcal{X}^n_b$, and refer to this suboptimal despeckler as BD-QMAP$_b$. The corresponding optimization problem is expressed as:
$$
    \argmin_{u^n \in \mathcal{X}^n_b} \frac{1}{n} \sum_{i=k}^n \left(\log u_i^2 + \frac{y_i^2}{u_i^2} + \frac{\lambda}{b} w_{u_{i-k+1}^i}\right), \label{eq:qmap-b-despeckle}
$$
where the term $c_{\mathbf{w}}(u^n)$ is scaled by $(n-k+1)/n$. This scaling has no impact on the optimal solution, as it can be absorbed into the hyperparameter $\lambda$. 

The negative log-likelihood term in $ \eqref{eq:qmap-b-despeckle} $ is separable across the $n$ observations, with each term depending only on $u_i$. Consequently, the Viterbi algorithm \cite{forney1973viterbi} can be used to efficiently solve this optimization problem. 

The Viterbi algorithm operates on a trellis diagram with $n$ stages, each containing $|\mathcal{X}_b|^k$ states. A state $s_i$ in the diagram corresponds to the subsequence $u_{i-k+1}^i$ of length $k$ in the candidate solution. The cost associated with state $s_i = u^i_{i-k+1}$ is updated recursively as:
$$
    C_i(s_i) = \min_{s' \in \mathcal{S}(s_i)} \left(\log u_i^2 + \frac{y_i^2}{u_i^2} + \frac{\lambda}{b} w_{u^i_{i-k+1}} + C_{i-1}(s')\right),
$$
for $i \in \{k, \dots, n\}$, where $\mathcal{S}(s_i)$ denotes the set of all states that transition into $s_i$, and $|\mathcal{S}(s_i)| = |\mathcal{X}_b|$. The state $s' = u_{i-k}^{i-1}$ represents the preceding state leading to $s_i$. The BD-QMAP$_b$ solution is obtained by minimizing $C_n(s_n)$ recursively.

\subsection{Theoretical Lower Bound}\label{subsec:experiment-theoretical-lb}

In \Cref{thm:MSE-lower-bound}, we derived a theoretical lower bound on the minimum achievable reconstruction error for piecewise-constant sources. To explore this bound empirically, we compare it with the performance of an ideal despeckler referred to as the \emph{genie-aided maximum likelihood (ML)} despeckler. This ideal estimator assumes access to the exact jump locations in the ground truth signal in addition to the noisy observations. With this side information, the reconstruction error is minimized by applying \Cref{eq:constant-estimator} to each constant segment between consecutive jump locations.
We evaluate this estimator and compare it to the lower bound, along with the experiments on piecewise constant source despeckling using BD-QMAP, as explained in the next section.

\subsection{Comparative analysis of despeckling for piecewise constant sources}\label{subsec:piecewise-comparison}
\begin{figure}
    \centering
    \includegraphics[width=0.7\linewidth]{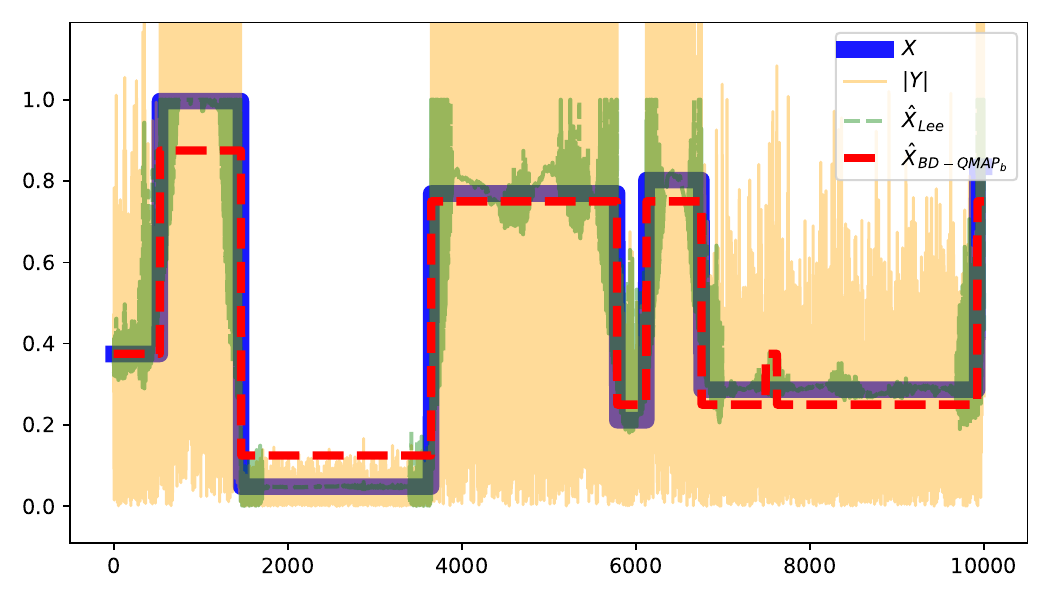}
    \caption{Piecewise constant source ($X$) with parameter $q_0=0.001$ sampled under speckle noise ($Y$), enhanced Lee and BD-QMAP$_b$ despeckled reconstructions, as discussed in \Cref{sec:experiments}.}
    \label{fig:source-demonstration}
\end{figure}

This subsection evaluates the effectiveness of BD-QMAP in despeckling piecewise-constant sources.  The results are compared against several well-known despeckling methods, including linear adaptive filters and enhanced versions thereof. Additionally, the role of regularization and quantization in BD-QMAP is explored to understand its impact on reconstruction quality and computational efficiency.

For numerical experiments with the piecewise-constant source introduced in \Cref{sec:piecewise}, we assume the absolutely continuous distribution is uniform over the interval $[0,1]$. A sample realization of this source under speckle noise is shown in \Cref{fig:source-demonstration}. For this source, the pivotal parameter affecting despeckler's performance is $q_0$, the probability of a signal level change, which determines the degree of structure in the source \cite{jalali2017universal}. To train the weights of \Cref{eq:qmap-weights} used in the BD-QMAP regularizer, we sample the random process for $10^7$ points with fixed $q_0$ and compute the empirical second-order distribution by counting occurrences of $b$-bit quantized sample pairs.

The BD-QMAP method offers two key practical advantages over existing despecklers. First, unlike methods such as \cite{aubert2008variational, shi2008nonlinear} and \cite{lopes1990adaptive}, BD-QMAP does not assume an upper bound on speckle power levels in the observations, making it robust to high-noise regimes where other filters fail or bypass the input. Second, BD-QMAP is flexible and does not require restrictive assumptions on the signal’s prior distribution, in contrast to Bayesian filters like those in \cite{kuan1985adaptive, lopes1990maximum, lee1980digital}. These features make BD-QMAP well-suited to a wide range of structured sources. \Cref{tab:performance-compare} summarizes the despeckling performance for test datasets of size 100, each containing piecewise-constant signals of length $n = 10^5$.

\subsubsection{BD-QMAP implementation}\label{subsubsec:bdqmap-piecewise-implement}

\begin{figure}
    \centering
    \includegraphics[width=0.7\linewidth]{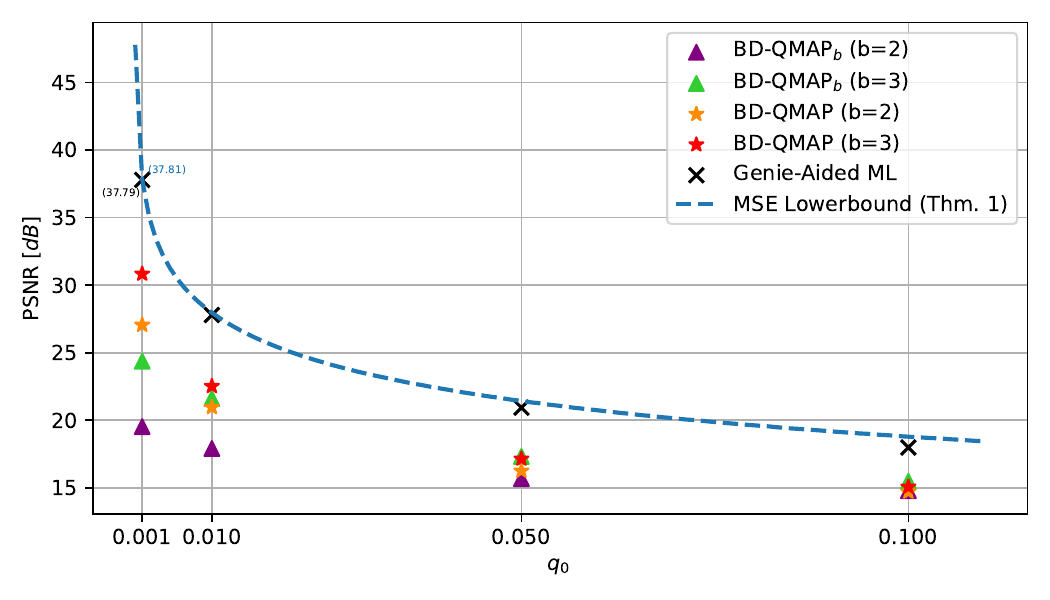}
    \caption{MSE lower bound of the genie-aided ML despeckler as derived in \Cref{thm:MSE-lower-bound} and the proposed BD-QMAP despecker for piecewise constant source with parameter $q_0$. ($\operatorname{PSNR}=10\log_{10}(\operatorname{MAX}^2/\operatorname{MSE})$ where $\operatorname{MAX}$ is the maximum value of signal's support set)}
    \label{fig:mse-upperbound}
\end{figure}

Using the Viterbi implementation described in \Cref{subsec:alg-considerations} for $k=2$, we efficiently implement BD-QMAP$_b$. As noted in \Cref{rmk:opt-challenge}, the fidelity term in BD-QMAP is not restricted to the quantized space. Therefore, we refine the BD-QMAP$_b$ solution by identifying detected jump locations and applying ML estimation on the intervals between jumps using \Cref{eq:constant-estimator}. This updated estimation is called BD-QMAP in our experiments. This refinement significantly improves performance, particularly for highly structured sources (i.e., small $q_0$). As shown in \Cref{tab:performance-compare}, even a low-complexity implementation with $b=2$ effectively detects jump locations, and the refinement step yields substantial performance gains. In \Cref{fig:mse-upperbound} the performance of BD-QMAP$_b$ and its refined version, BD-QMAP, is compared to the theoretical lower bound from \Cref{thm:MSE-lower-bound} and the genie-aided ML estimator from \Cref{subsec:experiment-theoretical-lb}.


\subsubsection{Regularization and quantization in BD-QMAP} 
\begin{figure}
    \centering
    \includegraphics[width=\linewidth]{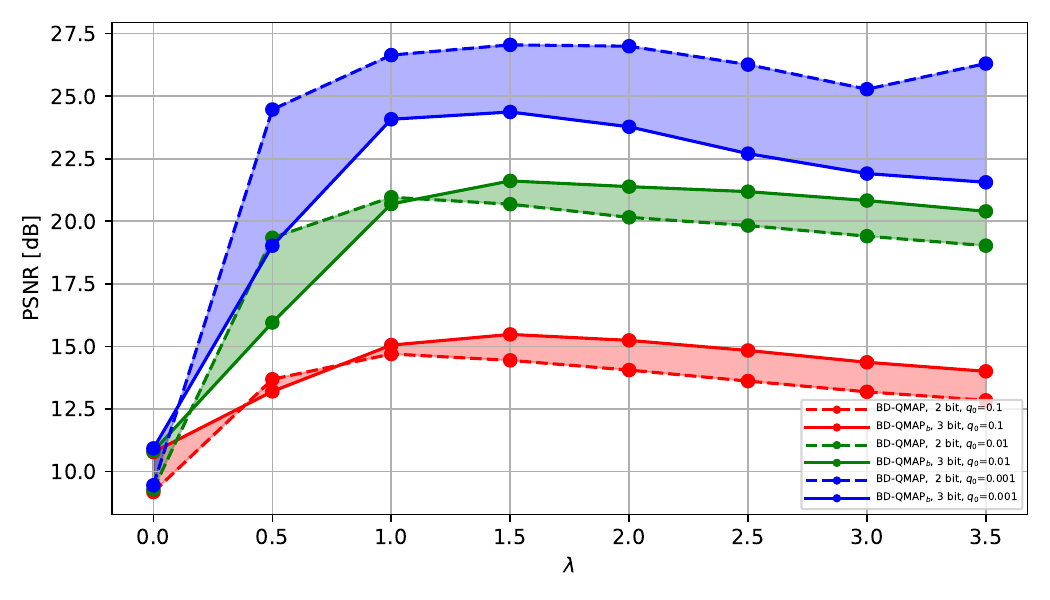}
    \caption{Effect of hyperparameter $\lambda$ in BD-QMAP$_b$ and BD-QMAP optimization for different choices of bits ($b$) and source structure ($q_0$).}
    \label{fig:qmap-lambda-dependence}
\end{figure}

To evaluate the impact of the BD-QMAP regularizer, we conduct experiments across a range of values for the hyperparameter $\lambda$ in \Cref{eq:qmap-despeckle}, as shown in \Cref{fig:qmap-lambda-dependence}. When $\lambda=0$, the despeckler fails to distinguish between sources with different levels of structure (as measured by $q_0$). Introducing a small regularization allows the denoiser to recover the empirical distribution of the ground truth in the reconstruction. However, excessively large values of $\lambda$ reduce fidelity to the noisy signal.

The number of bits used for quantization ($b$) also significantly impacts performance and computational complexity. While larger $b$ generally yields better reconstructions, for highly structured sources (e.g., high-resolution images), even small $b$ can effectively capture source structure. As demonstrated in \Cref{fig:qmap-lambda-dependence}, for $q_0 = 0.001$, the suboptimal BD-QMAP$_b$ solution with just $b=2$ bits captures the source structure well. After refinement, it achieves higher PSNR than BD-QMAP$_b$ with $b=3$, suggesting that smaller $b$ can be sufficient for highly structured data while reducing computational costs.


\begin{table}
\caption{Despeckling Performance Comparison (PSNR) for Piecewise Constant Source. 
}
\label{tab:performance-compare}
\centering
\renewcommand{\arraystretch}{1.25}
\begin{tabular}{lccc}
& $q_0=0.1$ & $q_0=0.01$ & $q_0=0.001$ \\ \cline{2-4} 
{Speckled Source} &8.710&8.742&9.055\\\hline\hline
{Box Car Filter} &13.226&16.321&16.976\\\hline
{Frost Filter \cite{frost1982model}} &12.923&13.900&14.226\\\hline
{Total Variation \cite{shi2008nonlinear}} &9.339&10.895&11.483\\\hline
{Lee Filter \cite{lee1980digital} / Enhanced \cite{lopes1990adaptive}} &10.505 / 13.865&18.402 / 19.703&22.303 / 22.577\\\hline
{Kuan Filter \cite{kuan1985adaptive}} / {Enhanced \cite{lopes1990adaptive}} &11.618 / 14.041&19.650 / 20.375& 23.017 / 23.244\\\hline
{BD-QMAP$_b$ (b=2 / b=3)} &14.786 / \textbf{15.478}& 17.909 / 21.613& 19.530 / 24.363 \\\hline
{BD-QMAP (b=2 / b=3)} &14.696 / 15.072&20.961 / \textbf{22.518}&27.052 / \textbf{30.831}  
\end{tabular}
\end{table}

\subsubsection{Linear adaptive filters}

All linear adaptive filters in \Cref{tab:performance-compare} require a window-size hyperparameter to define the region over which local statistics are computed. In our experiments, this value is set to $1 / 2q_0$, corresponding to half the expected interval length of constant signal pieces. This choice ensures the filters adapt to the level of structure in the source.

We also experiment with enhanced versions of these filters, incorporating heterogeneity adjustments as proposed in \cite{lopes1990adaptive}. This approach classifies regions as homogeneous, heterogeneous, or strongly speckled based on the local coefficient of variation of the observations, defined as $C_Y = \operatorname{Var}(Y) / \E[Y]$.

The enhanced filter operates as follows: If $C_Y$ falls below the noise coefficient of variation $C_W = \operatorname{Var}(W) / \E[W]$, the filter outputs the mean value of the local window. If $C_Y \geq C_{\max}$, where $C_{\max} = \sqrt{3}C_W$, the filter is bypassed, and the observed pixel is retained. Otherwise, the filter is applied normally. The enhanced versions of the filters in \Cref{tab:performance-compare} incorporate these modifications to improve despeckling performance.

\section{Conclusion}
A novel Bayesian despeckling method, BD-QMAP, is proposed for structured sources, providing a theoretically grounded approach to addressing speckle noise. Its performance is analyzed for piecewise-constant structured sources, with a derived lower bound on the minimum achievable MSE. Experimental results highlight the effectiveness of BD-QMAP in despeckling piecewise-constant sources, achieving state-of-the-art performance.

However, the theoretical understanding of despeckling and its fundamental limits remains significantly less developed compared to denoising in additive noise models, presenting substantial opportunities for further research. Additionally, extending the application of QMAP-based methods to image despeckling is a promising direction, which we plan to explore in future work.
\section*{Acknowledgment}

AZ and SJ  were supported by NSF grant CCF-2237538.  

\bibliographystyle{IEEEtran}
\bibliography{IEEEabrv, refs}

\end{document}

%% file: defns.tex
\usepackage{xspace}
\usepackage{bbm}
\def\diag{\mathop{\rm Diag}\nolimits}%
\newcommand{\Ac}{\mathcal{A}}

\newcommand{\Ec}{\mathcal{E}}

\newcommand{\Ic}{\mathcal{I}}

\newcommand{\Nc}{\mathcal{N}}

\newcommand{\Uc}{\mathcal{U}}

\newcommand{\Xc}{\mathcal{X}}

\newcommand{\ex}{{\rm e}}

\newcommand{\wv}{{\bf w}}

\newcommand{\nv}{{\bf n}}

\newcommand{\Xh}{{\hat{X}}}

\newcommand{\xh}{{\hat{x}}}

\let\E\relax
\DeclareMathOperator\E{E}
\let\P\relax
\DeclareMathOperator\P{P}

\def\textiid{i.i.d.\@\xspace}
\newcommand\iid{\ifmmode\text{ i.i.d. } \else \textiid \fi}

